**Minimal Cylinder Analysis Reveals the Mechanical Properties of Oncogenic Nucleosomes**

Mary Pitman [1,2], Yamini Dalal[1]* and Garegin A. Papoian[2]*

[1]Laboratory of Receptor Biology and Gene Expression

Center for Cancer Research

National Cancer Institute, NIH,

Bethesda MD

[2] Department of Chemistry and Biochemistry

Institute for Physical Science and Technology

University of Maryland

College Park, MD

*email: gpapoian@umd.edu; dalaly@mail.nih.gov

**Abstract**

Histone variants regulate replication, transcription, DNA damage repair, and chromosome segregation. Though widely accepted as a paradigm, it has not been rigorously demonstrated that histone variants encode unique mechanical properties. Here, we present a new theoretical approach called Minimal Cylinder Analysis (MCA) to determine the Young's modulus of nucleosomes from all-atom Molecular Dynamics (MD) simulations. Recently, we validated this computational analysis against *in vitro* single-molecule nanoindentation of histone variant nucleosomes. In this report, we further extend MCA to study the biophysical properties of hybrid nucleosomes that are known to exist in human cancer cells and contain H3 histone variants CENP-A and H3.3. We investigate the mechanism by which the elasticity of these heterotypic nucleosomes augments cryptic binding surfaces. Further, we derive biological predictions that might arise when such heterotypic nucleosomes take over large parts of the genome.

**Statement of Significance**

Nucleosomes are the base unit of eukaryotic genome organization. Histone variants create unique local chromatin domains that fine-tune transcription, replication, DNA damage repair, and faithful chromosome segregation. It is becoming increasingly clear that the mechanical response of chromatin, through material properties such as elasticity, regulates genetic function. We developed a theoretical method, validated by *in vitro* nanoindentation studies, called Minimal Cylinder Analysis (MCA), to determine the Young's modulus of nucleosomes from Molecular Dynamics simulations. We then postulate specific biological predictions about oncogenic hybrid nucleosomes using MCA. In the future, this computational method can serve as a fast and high-throughput tool to discern how macromolecular systems respond to mechanical forces.

1. **Introduction**

It is becoming increasingly clear that the elastic properties of chromatin regulates genetic function (1). Early evidence of the elastic behavior of chromatin came from the classic micromanipulation experiments of grasshopper chromosomes (2). Several subsequent studies have shown that chromatin acts as an elastic medium and that its constituent linker DNA behaves as an entropically driven elastomer (3, 4). Such studies on the physics of chromatin have led to new biological insight. For example, the pericentromere can act as a mechanical spring, governing chromosome separation and spindle length during mitosis (5). Thus, chromatin physics can help to address questions about chromatin ordering (6–8), how DNA is both stable and distortable (9, 10), how glassy DNA dynamics gives rise to cell-to-cell variability (11), and even how the mechanical micro-environment tunes genetic expression (12, 13). Chromatin states are altered by posttranslational modifications (PTMs) (14) and by histone variant deposition at the macromolecular scale (9). Consequently, the additive effects of nanoscale modifications are an essential component of chromatin chemical signaling pathways and may alter mechanical response.

In a recent study, we measured the material properties of CENP-A nucleosomes and binding partners, located primarily at centromeres, and H3 nucleosomes found throughout the chromosome arms (Melters, Pitman, Rakshit *et al. in press, Oct 2019*). Previously, we elucidated the structural mechanism underlying centromeric nucleosome plasticity *in silico* (15). To measure the Young's modulus of these nucleosomes and to understand the initial effects of kinetochore formation, we performed *in vitro* atomic force microscopy studies and compared these findings to *in silico* measurements (Melters, Pitman, Rakshit *et al. in press, Oct 2019*). This report explains in detail the novel methodology we developed for computational dissection of nucleoprotein complexes. Furthermore, we apply this method to investigate the material properties of other types of nucleoprotein complexes found in cancer cells, which are challenging to study experimentally due to their low abundance *in vivo*, and whose function appears to be detrimental to chromosomes. In many aggressive forms of cancer, CENP-A, a centromeric histone H3 variant, is overexpressed (16–18). Studies have demonstrated that either in cancer cells derived from patient tumors, or when artificially over-expressed, excess CENP-A is deposited outside the centromere, and stably retained there in the form of unexpectedly stable (19) hybrid nucleosomes containing CENP-A and H3.3 (17, 18, 20). In this report, we present our new computational approach and apply it to explore the material properties and biological impacts of hybrid CENP-A:H3.3 nucleosomes in cancer cells.

There exist several theoretical approaches to describe the elastic properties of macromolecules computationally. One such method is finite element analysis (FEA), where a mesh network models the structure, and energy is minimized in response to deformation (21, 22). However, the accuracy of this method requires system-specific parameterization to account for atomic interactions such as Coulombic forces. FEA at the nanoscale has produced results consistent with Molecular Dynamics (MD) when informed by atomistic simulation (23), but FEA lacks the built-in portability and resolution of MD. To achieve all-atom resolution, force-probe MD simulations have been implemented (24, 25). However, large systems such as macromolecular complexes are computationally costly, and unphysical force loading rates are typically required due to short simulation time. Lastly, coarse-grained MD force-fields have also been developed that are, excitedly, able to study the non-elastic deformation and fracture of macromolecules to simulate nanoindentation (26). The longer timescales achieved by coarse-grained methods are promising, but they lack the resolution of all-atom and may not resolve differences due to PTMs or variants. In the methodology we present here, we analyze all-atom resolution simulations of nucleosomes at extended time scales, and then use surface fluctuations to derive the modulus of elasticity in the absence of applied forces. The strength of our methodology is that it does not require extensive and expensive computational resources beyond equilibrated simulations.

The elastic moduli is derived by connecting equilibrium strain fluctuations with stress response (27). We employ this logic to obtain the elasticity of nucleosomes without applying an external force. Furthermore, we have introduced a simple temporal hierarchy when implementing our algorithm: first, the equilibrium trajectory is averaged over short timescale windows, and the resulting structures snuggly fit into encompassing cylindrical bounding domains. Afterward, the sequence of fluctuating cylinders is analyzed using solid mechanics, while also estimating the energy of the corresponding low-frequency vibrational mode from the equipartition theorem. Overall, our algorithm produces the absolute values of nucleosomes' Young's moduli without tunable parameters.

## 2. Methods

The goal of our analysis method is to calculate the Young's modulus of nucleosomes in the absence of applied forces. Essentially, this technique connects structural fluctuations observed in unbiased MD simulations, with the nucleosome's mechanical response, ultimately producing the absolute value of the Young's modulus. To analyze all-atom simulation data in such a way, we first treat the nucleosomes as mechanically homogenous elastic cylinders vibrating in a thermal bath. Next, we calculate the dimensions and fluctuations of what we term "minimal" cylinders over the ensemble of each trajectory. We define the cylinder dimensions as the minimum volume that contains the rigid exterior surfaces of the nucleosome.

To develop a simplified model for elasticity calculations, we make assumptions based on the known physical properties of nucleosomes. First, we apply an averaging technique to all-atom simulation data using continuum mechanics. Elastic continuum theory has been shown to predict material properties on the nanoscale when compared to experiments and analytical predictions (28, 29). We further reduce degrees of freedom and variability by utilizing the pseudo-symmetry and geometry of nucleosomes to treat them as homogenous circular cylinders. Next, we make simplifications on the mode of deformation studied. To compare to single-molecule nanoindentation, we assume that nucleosomes are compressed perpendicular to the axis of the cylinder. Therefore, we model nucleosome fluctuations as compression and expansion in the absence of shearing motions and attribute to this mode an equipartition of energy.

The workflow we used to determine the Young's modulus of nucleosomes, after obtaining an atomistic trajectory, is as follows:

2.1.1. Define the all-atom nucleosome coordinate system

2.1.2. Probe for rigid external cylinder bases and lateral surfaces

2.1.3. Retrieve average cylinder dimensions and variances

The output of these steps is then used to calculate the Young's modulus. Next, we will describe in detail each of these steps.

*2.1.1. Define the all-atom nucleosome coordinate system*

Analogous to the requirement of consistent orientation of nucleosomes in nanoindentation studies, we must first choose a standard nucleosome orientation. The question asked is this: if nucleosomes were to lie 'flat' on a surface, what would this orientation be? The alignment chosen alters the cylinder dimensions we will measure. Since we constrain our analysis to right circular cylinders, if the nucleosomes are tilted, the measured heights will be artificially large. To define our coordinate system, we first computed the principal axes of rotation and oriented the cylinder base to the plane of the first two principal axes (Fig. 1A) (Supporting Material).

*2.1.2. Probe for rigid external cylinder bases and lateral surfaces*

Since we are measuring elasticity without an applied force, we can consider the thought experiment: if one were to, hypothetically, push down on the nucleosome surface, at what point would compression become strongly hindered? To answer this, we need to find at what depth in the nucleosome residues would be contacted and resist further deformation. For example, an intrinsically disordered region or loop would easily collapse in response to force. In contrast, a structured α-helix of β-strand would resist deformation. Thus, we need a metric for local stiffness. The rigidity of residues can be quantified by the root mean square fluctuations, RMSF, of each residue throughout the simulation. High RMSF values correspond to increased fluctuation or decreased stiffness. Since RMSF is a time-averaged parameter, multiple time steps are required to calculate fluctuations of residues. Therefore, we divide the 1 μs simulation (from 0.6-1 μs) into 800 segments per simulation and output each residue's RMSF per segment. We probed how sensitive our analysis is to these trajectory-parsing parameters and found that they are robust through their Sensitivity Index (Supporting Material).

*2.1.3. Retrieve average cylinder dimensions and variances*

We next sort the C-α atoms by z-axis coordinates and select the z coordinate of the residue such that five rigid residues below an RMSF threshold are excluded outside the cylinder bounds, to find the height, h. We repeat this process for the r-axis to calculate the radius, r, of the cylinder per segment. For the cylinder distributions, we find $z_{avg}$ and $r_{avg}$; the variance or spread of the distributions; and the standard deviations, $\Delta r$ and $\Delta z$. This data was then used in our derivation for the Young's modulus as described below. We have made our code for calculating - the Young's modulus available an open-source tool (30).

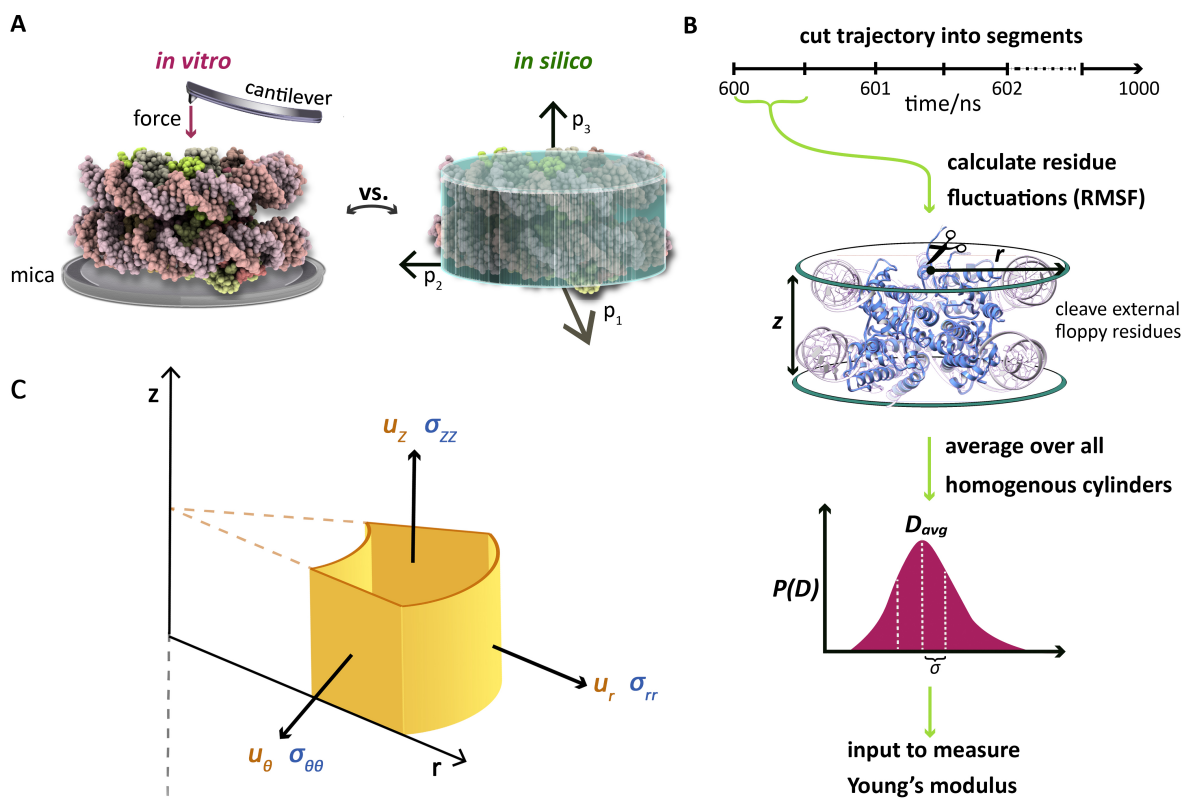

Figure 1: (A) Schematic that compares *in vitro* AFM single-molecule nanoindentation force spectroscopy, left, to our *in silico* modeling and analysis, right. In AFM, the applied force from the cantilever is normal to the mica surface. For our computational analysis, nucleosomes were oriented by the principal axes of the moment of inertia and then modeled as homogenous elastic cylinders. (B) The workflow for calculating Young's modulus *in silico*. Residue RMSF is calculated for each segment of the simulation to obtain an ensemble of cylinders. The average dimensions, $D_{avg}$, of the radius and height (r, z) and the standard deviation, σ, are then input to calculate Young's modulus. (C) A diagram to show the orientation of cylinders in relation to variables introduced in Eq. 2 and Eq. 4. Displacements ($u_r$, $u_\theta$, $u_z$) are shown in yellow. Stresses in the i-th direction from forces applied in the j-th direction, $\sigma_{ij}$, are shown in blue.

*2.2. All-atom computational modeling*

The software suite GROMACS 5.0.4 (31) was used to perform all-atom MD simulations. The force fields used were amber99SB*-ILDN (32, 33) for proteins, amber99SB parmbsc0 (34) for DNA, ions08 (35) for ions, and the TIP3P water model. Two nucleosome systems were built for simulation. First, the heterotypic CENP-A:H3.3 nucleosome was built from the crystallographic structure PDB ID: 3WTP (19). Subsequently, the CENP-A nucleosome was built with PDB ID: 3AN2 (36). The unresolved residues, CENP-A' 79-83, from the crystal structure (3AN2) were inserted using MODELLER (37). For energy minimization of the inserted residues, the N- and C-terminus were unconstrained. The 146 base pair α-satellite DNA, PDB ID: 3WTP (19), was aligned to all systems, as a control, using the CE algorithm (38) of PyMOL (39).

In order to assign charges of residues at a biological pH of 7.4, the GROMACS tool pdb2gmx was used. The boundaries of the simulation were set to a cuboid box with a minimum distance of 1.5 nm from the nucleosome with periodic boundary conditions. Na+ and Cl- ions were introduced to model an ionic physiological concentration of 150 mM NaCl. The Particle Mesh Ewald method was used for electrostatics with the Verlet cutoff scheme. Coulombic and Van der Waals potentials were used for non-bonded interactions with a cutoff distance at 1.0 nm. The LINCS algorithm was used to constrain hydrogen bonds.

Energy minimization was performed using steepest descent to a maximum energy of 100 kJ/mol. Following this, equilibration of the structure was carried out. The systems were heated to 300 K for 2000 ps with a DNA position restraint of $K = 1000$ kJ mol$^{-1}$ nm$^{-2}$ in the Canonical ensemble. Following this, thermal equilibration was performed for both DNA and protein at 300K for 2000 ps with position restraints defined as $K_{het} = 2.1e-5$ kJ mol$^{-1}$ nm$^{-2}$ assigned to the heterotypic nucleosome and $K_{cpa} = 2.5e-5$ kJ mol$^{-1}$ nm$^{-2}$ for the CENP-A nucleosome. These weak position constraints vary based on atom number in each simulation and restrain nucleosome rotations. Finally, the pressure was equilibrated for 1500 ps in the Isothermal-isobaric ensemble at 300 K, 1.0 bar pressure, and the position constraints $K_{het}$ or $K_{cpa}$.

Each production simulation was run for 1 microsecond at 300 Kelvin. Simulation temperature was V-rescaled using the modified Berendsen thermostat (40) with time constant 1.0 ps. The Parrinello-Rahman barostat (41) was used for pressure regulation at 1.0 bar, time constant of 2.0 ps. A simulation time step of 2 femtoseconds was used and coordinates, velocities, and energies saved every 2 ps. The non-bonded neighbors lists was updated at intervals of 20 femtoseconds. In order to analyze equilibrated sections of the production runs, the first 600 nanoseconds were not included in analysis. Detailed methods on all-atom structural analysis are provided in the Supporting Material.

## 3. Results

### 3.1. Sketch of the derivation of Young's modulus from MCA

We will present here main highlights from the derivation for Young's modulus from MCA. For those interested, a full, extended derivation is also included (Supporting Material). The work done in the deformation of an elastic material is stored in the form of strain energy, U. The strain energy density, u, the energy stored in small volume elements, can be useful to describe variable strains along a body that sum to the total strain energy:

$$U = \iiint_R u(r, \theta, z) r \, dr \, d\theta \, dz \, . \quad (1)$$

Because the extent of cylinder fluctuations is relatively small, in the range of 1.8 to 9.1% of the average radial or lateral dimension, we rely on linear elasticity and small-deformations' theory. Under these conditions, the strain energy density in cylindrical coordinates can be calculated for low magnitude stresses from arbitrary directions (42) as

$$u = \frac{1}{2}(\sigma_{rr}\varepsilon_{rr} + \sigma_{\theta\theta}\varepsilon_{\theta\theta} + \sigma_{zz}\varepsilon_{zz}) + (\sigma_{r\theta}\varepsilon_{r\theta} + \sigma_{\theta z}\varepsilon_{\theta z} + \sigma_{zr}\varepsilon_{zr}) = \frac{1}{2}Tr(\boldsymbol{\sigma\varepsilon}) \text{, (2)}$$

where $\sigma_{ij}$ is the stress in the i-th direction from force applied in the j-th direction and $\varepsilon_{ij}$ is the strain in the i-j plane (Fig. 1C). Further explanation for the form of Eq. 2 in cylindrical coordinates is provided in section S4 where, briefly, we apply the cyclic property of trace on the second order symmetric tensors, stress and strain, to arrive at Eq. 2 (Supporting Material). In the absence of shear stresses and using Hooke's law, the strain energy density in Eq. 2 can also be written in the form

$$u = \frac{\nu\mu}{1-2\nu}(\varepsilon_{rr} + \varepsilon_{\theta\theta} + \varepsilon_{zz})^2 + \mu(\varepsilon_{rr}^2 + \varepsilon_{\theta\theta}^2 + \varepsilon_{zz}^2) \text{, (3)}$$

where $\mu$ is the shear modulus and is related to Young's modulus, E, by $\mu = E / 2(1+ \nu)$ and $\nu$ is the Poisson ratio (42). For displacements ($u_r$, $u_\theta$, $u_z$) in cylindrical coordinates (r, θ, z) as shown in Fig. 1C (43):

$$\varepsilon_{rr} = \frac{\partial u_r}{\partial r}, \quad \varepsilon_{\theta\theta} = \frac{u_r}{r} + \frac{1}{r}\frac{\partial u_\theta}{\partial \theta}, \quad \varepsilon_{zz} = \frac{\partial u_z}{\partial z} \text{. (4)}$$

In the special case of a homogeneous axisymmetric cylinder where the center-of-mass is at the origin (Fig. 1C), $\frac{\partial u_\theta}{\partial \theta} = 0$ and at the walls of the cylinder $\frac{\partial u_r}{\partial r} = \frac{u_r}{r_{avg}}$, (6) which is $\frac{\Delta r}{r_{avg}}$ from MCA. Therefore, in this specific case, $\varepsilon_{rr} = \varepsilon_{\theta\theta}$ in Eq. 4. More detail on how we arrive at these conclusions is provided (Supporting Material).

To calculate the strain energy density in Eq. 3, we input the dependence of the shear modulus on E and relations found from Eq. 4 to obtain

$$u = \frac{E}{2(1+\nu)}\left[\frac{\nu(\varepsilon_{zz}+2\varepsilon_{rr})^2}{(1-2\nu)} + \varepsilon_{zz}^2 + 2\varepsilon_{rr}^2\right] \cdot (5)$$

Strain values in Eq. 5 are calculated from the measured quantities $r_{avg}$, $z_{avg}$, $\Delta r$, and $\Delta z$ from MCA (Methods 2.1.3). We next focus on the acoustic cylindrical mode of motion that describes compression in the z-axis along with radial extension (and *vice versa*). The average potential energy of this mode is estimated from the equipartition theorem, $U = \frac{1}{2}k_b T$, where $k_b$ is the Boltzmann constant and T is the simulation temperature, 300 K. We then integrate Eq. 5 over the body volume, Eq. 1, and with the above-mentioned energy from equipartition theorem, we solve for Young's modulus:

$$E = \frac{k_b T(1-\nu-2\nu^2)}{V(\varepsilon_{zz}^2 - \nu\varepsilon_{zz}^2 + 2\varepsilon_{rr}^2 + 4\nu\varepsilon_{zz}\varepsilon_{rr})} \cdot (6)$$

This final form for the derivation of Young's modulus, shown in Eq. 6 is one of the main findings of this work.

### 3.2. *Experimental validation of MCA model*

For our analyses from our previous work (Melters, Pitman, Rakshit *et al. in press, 2019*), we used the Hertz model with spherical indenter geometry for Young's Modulus measurements (44). The Hertz model assumes that the substrate is an isotropic, elastic solid and is valid for small indentations and low forces, in the linear regime. To check for elastic dependence on the point probed, we measured the Young's modulus across mononucleosomes and found that the effective elasticity is surprisingly homogenous across the surface (Melters, Pitman, Rakshit *et al. in press, 2019*). This finding is consistent with the model of MCA, which treats nucleosomes as homogenous elastic solids.

*3.3. Experimental validation for in silico Young's modulus calculations*

In our study of the mechanical properties of nucleosome variants on the chromatin fiber, we applied our *in silico* methodology to measure the Young's Modulus of nucleosomes (Melters, Pitman, Rakshit *et al. in press, 2019*). We measured the elasticity three systems: (1) the canonical nucleosome, H3, (2) the centromeric variant of H3, CENP-A, (3) and CENP-A nucleosomes bound to CENP-C. We also measured the elasticity of these substrates *in vitro* using single-molecule nanoindentation force spectroscopy. Our *in silico* algorithm to determine Young's modulus showed quantitative agreement with *in vitro* measurements (Fig. 2A). These nanoindentation studies provide robust validation for our model and indicate that our assumptions used to calculate elasticity are acceptable simplifications.

*3.4. Young's modulus of the hybrid CENP-A:H3.3 vs. the CENP-A nucleosome*

After validating MCA against *in vitro* single-molecule force studies previously (Melters, Rakshit, Pitman et al. *in press, 2019*), we extend this method to study the elastic properties of a heterotypic cancer-specific nucleosome in this work. This unique variant nucleosome has one copy of CENP-A and one copy of H3.3, and is enriched at CENP-A ectopic sites in chromatin (18), some of which are well documented fragile sites in the chromatin fiber (17). The heterotypic nucleosome was found to be surprisingly stable, regardless of the unique docking interface formed between two divergent H3 variants (19). What, then, causes the apparent stability?

To address this question, we computationally assessed the elastic properties of the heterotypic nucleosome. We discovered an intermediate Young's modulus of hybrid CENP-A:H3.3 nucleosomes (8.3±1.7 MPa) compared to CENP-A nucleosomes (6.9 ±0.8 MPa)*,* and canonical H3 nucleosomes (9.9 ± 0.6 MPa, Fig. 2A). This result contradicts the idea that unfavorable contacts may form between the CENP-A:H3.3 heterodimer and disrupt the stability of the hybrid nucleosome. Since our methodology uses an averaging technique over the structure of the nucleosome, we next asked how the dynamics of the heterotypic nucleosome gives rise to its intermediate elasticity. Two hypotheses were considered: first, the heterotypic nucleosome presents an averaged global shift in nucleosome dynamics; or secondly, there may be sequestered regions within the heterotypic nucleosome that display canonical or centromeric nucleosome dynamics.

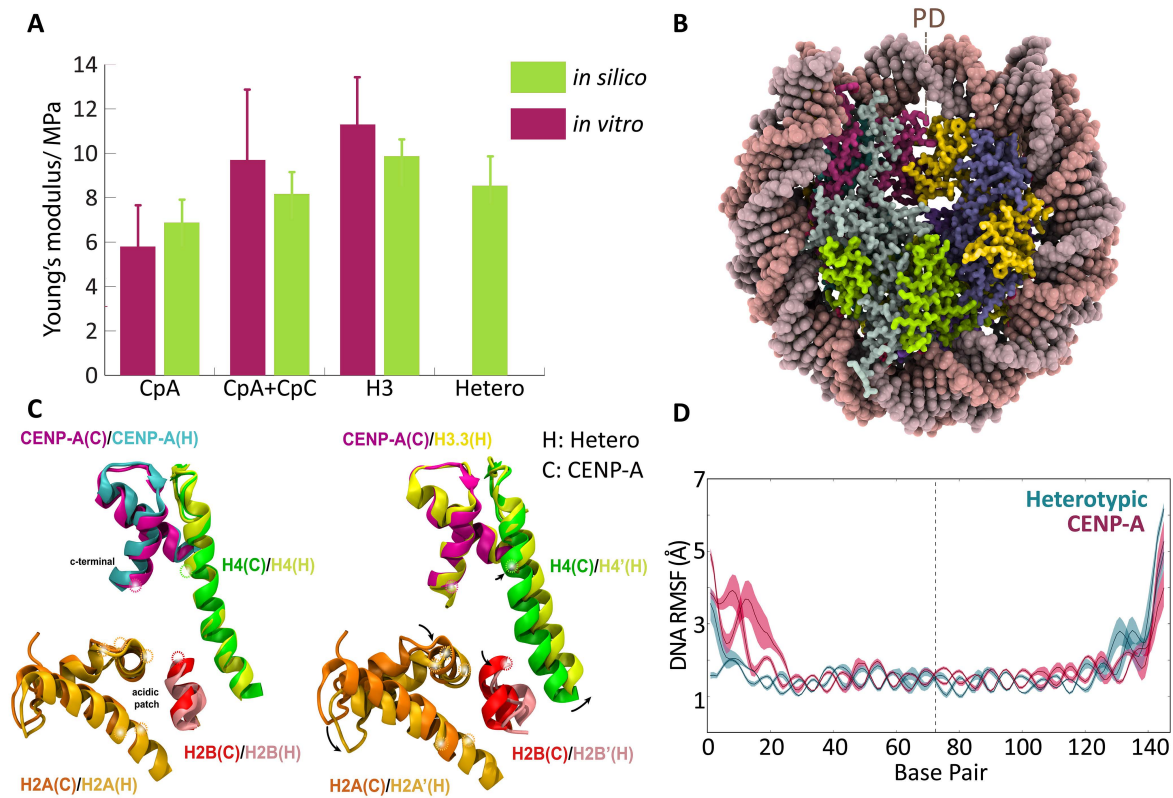

Figure 2: (A) Young's modulus of CENP-A nucleosomes, CpA; CENP-A bound to CENP-C, CpA+CpC; the canonical nucleosome, H3; and the CENP-A:H3.3 containing heterotypic nucleosome, hetero. AFM measurements *in vitro* are shown in magenta and *in silico*

measurements are shown in green. (B) The structure of the heterotypic nucleosome. Histones CENP-A are shown in magenta, H3.3 in yellow, H4 in dark slate-blue, H2A in light grey, and H2B in green. (C) The overlay of the CENP-C binding sites by minimum RMSD obtained from the representative structure of the first principal component. System "C" denotes histones from the CENP-A nucleosome and "H" from the CENP-A:H3.3 heterotypic nucleosome. CENP-C binding residues are highlighted. (D) Root-mean squared fluctuations (RMSF) of DNA of the CENP-A nucleosome, in magenta, and the heterotypic nucleosome, in blue.

*3.3 The rigidified heterotypic nucleosome is permissive to CENP-C binding*

The essential docking protein to initiate kinetochore formation is CENP-C, which binds to the surface of CENP-A (45). It is still unknown if the heterotypic CENP-A:H3.3 nucleosome is implicated in the formation of neocentromeres. Consistent with reduced flexibility compared to CENP-A (Fig. 2A), we found a more tightly bound four-helix bundle interface between H3.3 and CENP-A in the heterotypic nucleosome (Fig. S1A in the Supporting Material). CENP-C docks by interaction with the c-terminal tail of CENP-A in this region and binds across the nucleosome surface with the basic residues of H2A (residues 60, 63, 89-91 in drosophila melanogaster) and H2B (46).

Therefore, we analyzed the CENP-C binding platform to see if the heterotypic nucleosome is permissive to CENP-C. To do so, we performed Principal Component Analysis (PCA) and animated the first major mode of motion, the first Principal Component (PC1). Visualization of PC1 revealed that the CENP-A containing tetramer of the heterotypic nucleosome rocks apart less than the H3.3 tetramer and is more compact, similar to the CENP-A nucleosome (Movie S1). Indeed, the heterotypic nucleosome contains asymmetric and partitioned dynamics, where features of CENP-A nucleosome behavior are maintained. In PC1, we see that the CENP-C binding site from the CENP-A nucleosome is preserved in the heterotypic nucleosome (Fig. 2C, left). The c-terminus of H3.3, which does not bind CENP-C, is as a comparison extended away from the acidic patch in the heterotypic nucleosome, right (Fig. 2C, Fig. S1B). This analysis shows that the correct coordination of binding residues for CENP-C is maintained in the CENP-A facing side of the heterotypic nucleosome, making it permissive to the double arginine anchor mechanism of both the CENP-C central domain, R522, R525 (46), previously observed *in vitro* (19), and to the CENP-C motif, R717, R719 (14, 46).

We further discuss the dynamics of the heterotypic nucleosome and the exposure of the CENP-N binding site, the CENP-A RG loop (Supporting Material).

*3.4 DNA dynamics of histone variants is partitioned by the heterotypic nucleosome pseudo-dyad*

A fundamental question is if heterotypic nucleosomes are also able to bind linker histones (LHs) to form a chromatosome similarly to the canonical nucleosome (47, 48). The LH globular domain docks to the entry-exit sight of canonical nucleosomes (49), and the LH disordered tails bind to linker DNA, holding DNA ends together (50). A distinctive difference between canonical nucleosomes and CENP-A nucleosomes is the markedly lower affinity of the latter for LHs (51). Specifically, the LH H1 in unable to bind CENP-A nucleosomes due to a shorter CENP-A $\alpha$N helix in contrast to canonical nucleosomes, thus resulting in unstable entry and exit DNA strands (52). We then asked how the intermediate rigidity of the heterotypic nucleosome (Fig. 2A) affects DNA dynamics.

First, we calculated the root-mean-square fluctuation (RMSF) of DNA over three segments of our analyzed trajectories. We found that the presence of both CENP-A and H3.3 results in a symmetry breaking in DNA dynamics across the pseudo-dyad (Figure 2D). We observed increased DNA motion in the heterotypic nucleosome proximal to the CENP-A histone in contrast to the H3.3 histone. This region is of interest because the globular domain of H1 binds to the DNA minor groove on-dyad (52).

Furthermore, the asymmetry in DNA dynamics propagates to the DNA entry-exit ends. In the heterotypic nucleosome, we found increased DNA end fluctuations on the end proximal to CENP-A and decreased fluctuations proximal to H3.3 (Figure 2D). The disordered tails of H1 rely on the presence of DNA crossing at the entry and exit ends for nucleosome affinity and to compact the fiber (52, 53). Therefore, we next measured the DNA end-to-end distance in comparison to CENP-A nucleosomes. We found that the DNA ends of the heterotypic are closer together with a probability similar to that of canonical nucleosomes (Fig. S1D). More open configurations were least likely to occur by the heterotypic nucleosome Fig. S1D). The increased likelihood of close DNA end configurations suggests that heterotypic nucleosomes may bind LHs.

The principal finding from these simulation analyses is that the heterotypic nucleosome displays the dynamics of both canonical and centromeric nucleosomes, resulting in an overall intermediate elasticity. Our findings show that the presence of one H3.3 histone variant induces increased stability and that the CENP-A histone intrinsically induces a more elastic phenotype. These results provide a theoretical basis for experimental findings that reported on the surprising stability of the CENP-A:H3.3 containing nucleosome (19).

**Discussion**

When CENP-A is overexpressed in human cancer cells (17, 54, 55), CENP-A appears to take advantage of H3.3 chaperones, such as HIRA and DAXX (16, 18). The role of H3.3 chaperones in CENP-A deposition away from the centromere provides a logical pathway for the formation of hybrid CENP-A:H3.3 nucleosomes as dimer H3.3/H4 and CENP-A/H4 pairs may fortuitously co-assemble into tetramers on the DNA at regions of high turnover (56–58). Indeed, H3.3 chaperones are implicated in the ectopic formation of heterotypic nucleosomes in cells with increased survivability in the presence of DNA damage (18). The formation and retention of heterotypic nucleosomes on the chromatin fiber may be further augmented by our findings here that they are more rigid (8.3±1.7 MPa) than CENP-A alone (6.9±0.8 MPa) (Fig. 2A). In agreement with this hypothesis, CENP-A nucleosomes are more easily dissembled than H3 *in vitro* by NAP-1 or heparin destabilization (53), and by magnetic tweezers (59). The decreased flexibility of the heterotypic nucleosome may then make it a safe harbor for ectopically located CENP-A histones to evade eviction. Put another way, the enhanced rigidity of these particles might explain why they persist ectopically, when non-hybrid CENP-A nucleosomes may be easily removed by transcription or remodeling, were they to stochastically accumulate ectopically in normal cells (60).

Even more fascinating to consider is the dynamics of the heterotypic nucleosome, which is predicted to alter the accessibility of cryptic bindings sites and result in downstream biological effects. Our findings suggest that heterotypic nucleosomes are competent to bind CENP-C, the structural scaffold for inner kinetochore assembly (61). The CENP-A:H3.3 containing nucleosome binds the CENP-C central region *in vitro* (19) and ectopic mislocalization of CENP-A results in neocentomeres (16, 17, 20, 55, 62). However, the biological impact of these phenomena depends on the subsequent recruitment of proteins for microtubule attachment. Minimally, bound LHs or inner kinetochore proteins may further rigidify the heterotypic nucleosome and facilitate CENP-A retention ectopically.

**Conclusion**

The elasticity of nucleosomes has biological relevance due to the mechanical sensing of large macromolecules and histone variant-specific assemblies such as, in the case of CENP-A, CENP-C and the entire inner kinetochore complex. In the absence of irreversible distortions to the structure, where binding partners or nanomachines exert forces in the elastic range, our newly developed method, MCA, can be applied to measure Young's moduli of various nucleosome complexes that are of low abundance in cells of specific lineages. A significant benefit to this analytical method is that it can be applied to analyze equilibrium trajectories, enabling a computationally efficient and potentially high-throughput way to quantify elasticity and form testable predictions.

**Author Contributions**

Conceptualization: MP, GAP; Methodology: MP; Investigation: MP; Writing: MP, GAP, and YD.; Funding Acquisition: GAP, and YD; Visualization: MP; Supervision: GAP and YD.

**Acknowledgements**

We thank Carlos S. Floyd for independent validation of MCA code and feedback. For further comments, we thank Daniël P. Melters and Aravind Chandrasekaran. YD and MP are supported by the Intramural Research Program of the National Institutes of Health. GAP is supported by NSF grant CHE-1800418.

**Supporting Citations**

References (63–66) appear in the Supporting Material.

**Supporting Material**

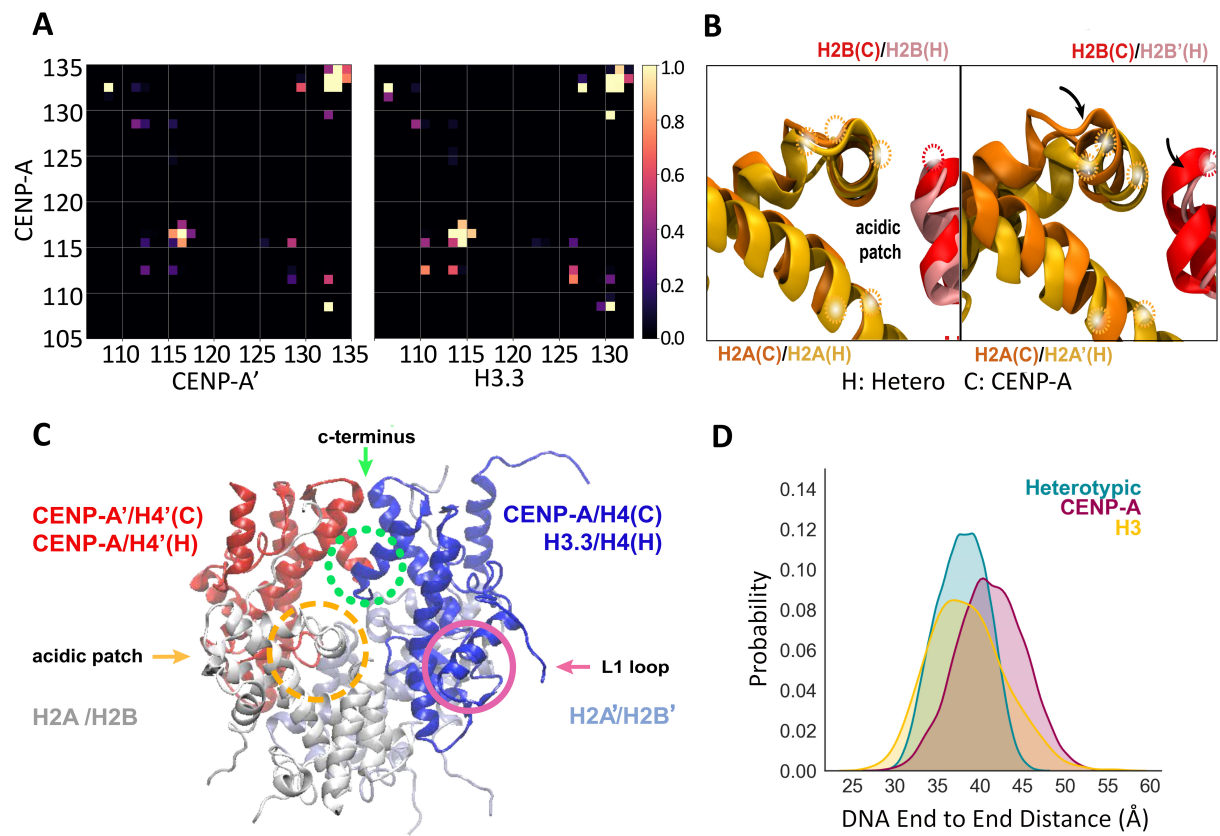

**Figure S1:** (A.) Contact analysis at the 4-helix bundle interface of CENP-A:CENP-A in the context of the CENP-A nucleosome on the left in comparison with CENP-A:H3.3 in the heterotypic nucleosome. Increased brightness of color shows the propensity of C-α contact within 8 Å. Black means that contact does not occur, and pale yellow indicates contact at all time-steps. (B.) The overlay of the acidic patches from the representative structure of the first principal component. System "C" denotes histones from the CENP-A nucleosome and "H" from the CENP-A:H3.3 heterotypic nucleosome. On the left, the CENP-A region from system C is shown with minimum RMSD alignment to the CENP-A region of the heterotypic nucleosome. For comparison, on the right, the H3.3 domain of H is compared to CENP-A in C. CENP-C

binding residues are highlighted. (C.) Representative image from Movie S1, which depicts histone core motions of the first principal component. Regions of interest in the PCA movie are highlighted such as the histones (where H is the Heterotypic nucleosome and C is the CENP-A nucleosome), the acidic patch in the dashed orange circle, the L1 loop in the pink circle, and the CENP-A or H3.3 c-terminal region in the dotted green circle. (D.) The histogram of DNA end to end distances for the Heterotypic nucleosome in teal, the CENP-A nucleosome in magenta, the H3 nucleosome in yellow.

**S1.** *Nucleosome Orientation*

The axes of symmetry of the nucleosomes align to the three principal axes. We confirmed that the axes were orthogonal and performed a rotational transformation of the atomic coordinates so that the principal axes aligned to the Cartesian coordinate system with the center-of-mass at the origin.

**S2.** *Parameter Sensitivity*

To calculate the Young's modulus, we rationally selected the input parameters that define how the trajectory is parsed (jump, cutoff, and threshold). As explained in greater detail previously (Methods), jump specifies the number of frames over which we calculate local flexibility through the time-averaged parameter, residue RMSF. Cutoff specifies the RMSF value below which residues are classified as rigid. Lastly, threshold defines the number of flexible residues cleaved from the exterior surfaces to determine the Minimal Cylinder. The input for each parameter is constant across systems.

We will first address the parameter sensitivity of our algorithm to jump and its selection. Since jump parses the trajectory into collections of N frames, jump should be a minimal value to produce RMSF data with a spread. If the value of jump is too small, there is unclear separation of rigid and flexible residues. However, if the trajectory is parsed too extensively, there is loss of sampling of cylinder dimensions. Therefore, we selected a jump value of 20 frames since this resulted in a distinguishable spread in RMSF data. We assessed jump values ranging from 20 to 80 frames and found from sensitivity analysis that to first-order jump is not a parameter does not tune Young's modulus systematically. The Sensitivity Index (SI), which measures the relative rate of change of the output to the relative rate of change of input, was found to be -0.004 with an $R^2$ of 0.0006, which shows a lack of linear correlation in the data. The standard deviation of the Young's modulus values when varying jump was found to be ± 0.76 MPa. To first-order, jump , therefore, needs to be selected to parse the trajectory and provide resolution to the RMSF values. However, our analysis is not sensitive to the selection of a specific jump value.

In order to select the cutoff parameter, the RMSF should be calculated for the unique system. We selected our RMSF cutoff value so that it falls within the range of RMSF data retrieved and is greater than the stable population of residues. We tested the range of RMSF data produced with values selected distinctly within the bounds of the RMSF extrema. We found that the SI to cutoff is -0.36 with an $R^2$ of 0.54 due to a drop in output values at the upper RMSF bounds. However, when a rigidity cutoff below 80% of the RMSF maxima is selected, we find reduced first-order sensitivity (SI = 0.044, $R^2$= 0.75). If values are selected that are greater than the RMSF maxima, no flexible residues will be cleaved from the analyzed dataset. Likewise, if values are selected below the minima, the algorithm will not retrieve cylinder dimensions.

The threshold value was selected so that exterior floppy residues are cleaved from dimension calculations, and boundary motions are probed versus internal motions. We selected a value of 10 residues to apply across our nucleosome systems and found that from a range of 5-12 residues cleaved, the algorithm produces a SI of 0.32, $R^2 = 0.47$ for threshold. To first-order MCA is most sensitive to threshold. To control for potential variability, all parameters were held constant during analysis.

Our analysis shows that these parameters do not influence the final order of magnitude of the Young's modulus and results in a final value within our calculated errors from three segments of the analyzed simulation. When held constant across systems, qualitative trends are recapitulated.

## S3. *All-atom Structural Analysis*

Simulation data was truncated to include the final 400 ns for analysis and then contact analysis was performed. A cutoff distance of 8 Å was used between histone Cα atoms to compare dimer interface formed between CENP-A:CENP-A vs. CENP-A:H3.3. A value of 1 indicates contact at all frames of the analyzed trajectory and a value of 0 indicates an absence of contact during simulation. Principle component analysis (PCA) was performed on the histone core based on previously published methods (1). The magnitude of motion is multiplied by a factor of 5 in the movies to amplify motions for visual clarity.

DNA was analyzed by residue RMSF with errors calculated over three trajectory segments from 600 ns to 1000 ns of total simulation time. DNA end distances were then calculated from the entry to exit ends of the heterotypic nucleosome and the CENP-A nucleosome, compared to H3 (1).

## S4. *Extended Derivation of Young's modulus from MCA*

The work done in the deformation of an elastic material is stored in the form of strain energy, U. The strain energy density, u, the energy stored in small volume elements, can be useful to describe variable strains along a body that sum to the total strain energy:

$$U = \iiint_R u(r, \theta, z) r \, dr \, d\theta \, dz \quad . \quad (1)$$

Because the extent of cylinder fluctuations is relatively small, in the range of 1.8 to 9.1% of the average radial or lateral dimension, we rely on linear elasticity and small-deformations' theory. Under these conditions, the strain energy density in cylindrical coordinates can be calculated for low magnitude stresses from arbitrary directions (2) as

$$u = \frac{1}{2}(\sigma_{rr}\varepsilon_{rr} + \sigma_{\theta\theta}\varepsilon_{\theta\theta} + \sigma_{zz}\varepsilon_{zz}) + (\sigma_{r\theta}\varepsilon_{r\theta} + \sigma_{\theta z}\varepsilon_{\theta z} + \sigma_{zr}\varepsilon_{zr}), \quad (2)$$

where $\sigma_{ij}$ is the stress in the i-th direction from force applied in the j-th direction and $\varepsilon_{ij}$ is the strain in the i-j plane (Fig. 1C). In Cartesian coordinates, the strain energy density of a volume element under arbitrary stresses can be found at Eq. 8.2.18 of (2) and is given as

$$u = \frac{1}{2}\left(\sigma_{xx}\varepsilon_{xx} + \sigma_{yy}\varepsilon_{yy} + \sigma_{zz}\varepsilon_{zz}\right) + \left(\sigma_{xy}\varepsilon_{xy} + \sigma_{yz}\varepsilon_{yz} + \sigma_{zx}\varepsilon_{zx}\right), \quad (S1)$$

where $\sigma_{ij}$ is the stress in the i-th direction from force applied in the j-th direction and $\varepsilon_{ij}$ is the strain in the i-j plane. In the main text, Eq. 2 can be derived from and is in the same form as (S1). In the absence of internal torques, stress and strain are both second order symmetric tensors (3, 4). This fact can then be used to understand the origin of Eq. S1. As a note, the derivation of Eq. 2 from Eq. S1 can also be done using trigonometric identities or Einstein summation notation. First,

$$\boldsymbol{\sigma\varepsilon} = \begin{bmatrix} \varepsilon_{xx} & \varepsilon_{xy} & \varepsilon_{xz} \\ \varepsilon_{xy} & \varepsilon_{yy} & \varepsilon_{yz} \\ \varepsilon_{xz} & \varepsilon_{yz} & \varepsilon_{zz} \end{bmatrix} \begin{bmatrix} \sigma_{xx} & \sigma_{xy} & \sigma_{xz} \\ \sigma_{xy} & \sigma_{yy} & \sigma_{yz} \\ \sigma_{xz} & \sigma_{yz} & \sigma_{zz} \end{bmatrix}. \quad (S2)$$

We will take the trace of the matrix product, and so the diagonal elements are

$$(\boldsymbol{\sigma\varepsilon})_{11} = \sigma_{xx}\varepsilon_{xx} + \sigma_{xy}\varepsilon_{xy} + \sigma_{zx}\varepsilon_{zx},$$

$$(\boldsymbol{\sigma\varepsilon})_{22} = \sigma_{xy}\varepsilon_{xy} + \sigma_{yy}\varepsilon_{yy} + \sigma_{zy}\varepsilon_{zy},$$

$$(\boldsymbol{\sigma\varepsilon})_{33} = \sigma_{xz}\varepsilon_{xz} + \sigma_{yz}\varepsilon_{yz} + \sigma_{zz}\varepsilon_{zz}. \quad \text{(S3)}$$

Therefore,

$$Tr(\boldsymbol{\sigma\varepsilon}) = (\sigma_{xx}\varepsilon_{xx} + \sigma_{yy}\varepsilon_{yy} + \sigma_{zz}\varepsilon_{zz}) + 2(\sigma_{xy}\varepsilon_{xy} + \sigma_{yz}\varepsilon_{yz} + \sigma_{zx}\varepsilon_{zx}), \quad \text{(S4)}$$

And from Eq. S1,

$$u = \frac{1}{2}Tr(\boldsymbol{\sigma\varepsilon}). \quad \text{(S5)}$$

The form shown in Eq. S5 becomes useful because of the cyclic property of trace. We will consider a transformation matrix, **O**, which is any orthonormal basis:

$$\frac{1}{2}Tr\big((\boldsymbol{O\sigma O^T})(\boldsymbol{O\varepsilon O^T})\big). \quad \text{(S6)}$$

Since **O** is any orthonormal basis, $\boldsymbol{O^T O} = \boldsymbol{1}$, Eq. S6 simplifies to

$$\frac{1}{2}Tr(\boldsymbol{O^T O \sigma \varepsilon}) = \frac{1}{2}Tr(\boldsymbol{\sigma\varepsilon}). \quad \text{(S7)}$$

This means that Eq. S1 can be written in the form Eq. S5, the form of Eq. S1 will be invariant to any orthonormal basis set. Therefore, Eq. S1 in Cartesian coordinates can be written in cylindrical coordinates as shown in Eq. 2.

From Eq. 2, since our output from the MCA is in terms of strain, we solve for strains from the stress-strain relations of Hooke's Law, which can be found from a Solid Mechanics reference at Eq. 6.1.8 (2). When solving for on diagonal stresses, the standard result is obtained that

$$\sigma_{rr} = \frac{E}{(1+v)(1-2v)}[(1-v)\varepsilon_{rr} + v(\varepsilon_{\theta\theta} + \varepsilon_{zz})],$$

$$\sigma_{\theta\theta} = \frac{E}{(1+v)(1-2v)}[(1-v)\varepsilon_{\theta\theta} + v(\varepsilon_{rr} + \varepsilon_{zz})],$$

$$\sigma_{zz} = \frac{E}{(1+\nu)(1-2\nu)}[(1-\nu)\varepsilon_{zz} + \nu(\varepsilon_{\theta\theta} + \varepsilon_{rr})] \quad (S8)$$

where E is Young's modulus and ν is the Poisson ratio (2). Using the relations found in Eq. S8, in the absence of shear stresses and using Hooke's law, the strain energy density in Eq. 2 can also be written in the form

$$u = \frac{\nu\mu}{1-2\nu}(\varepsilon_{rr} + \varepsilon_{\theta\theta} + \varepsilon_{zz})^2 + \mu(\varepsilon_{rr}^2 + \varepsilon_{\theta\theta}^2 + \varepsilon_{zz}^2) \quad (3)$$

where $\mu$ is the shear modulus and is related to Young's modulus, E, by $\mu = E/2(1+\nu)$. For displacements ($u_r$, $u_\theta$, $u_z$) in cylindrical coordinates (r, θ, z) as shown in Fig. 1C (5):

$$\varepsilon_{rr} = \frac{\partial u_r}{\partial r}, \quad \varepsilon_{\theta\theta} = \frac{u_r}{r} + \frac{1}{r}\frac{\partial u_\theta}{\partial \theta}, \quad \varepsilon_{zz} = \frac{\partial u_z}{\partial z} \quad (4)$$

In the case of a homogeneous axisymmetric cylinder, the center-of-mass at the origin (Fig. 1C), $\frac{\partial u_\theta}{\partial \theta} = 0$ and at the walls of the cylinder $\varepsilon_{rr} = \frac{u_r}{r_{avg}}$. For the case of a homogenous cylinder, we will use the notation Δr for the displacement in r, $u_r$. Therefore, in this specific case, $\varepsilon_{rr} = \varepsilon_{\theta\theta}$ in Eq. 4. For homogenous cylinders (6):

$$\varepsilon_{rr} = \frac{\partial u_r}{\partial r} = \left.\frac{\partial u_r}{\partial r}\right|_{r=r_{avg}} = \frac{\Delta r}{r_{avg}} \quad (S9)$$

where $\frac{\Delta r}{r_{avg}}$ is input from MCA. We will then use the relation from Eq. 4 that in our case $\varepsilon_{rr} = \varepsilon_{\theta\theta}$ and rewrite. Eq. 3 in terms of the Young's modulus, our solvable, and strains to obtain

$$u = \frac{\nu E}{2(1-2\nu)(1+\nu)}(2\varepsilon_{rr} + \varepsilon_{zz})^2 + \frac{E}{2(1+\nu)}(2\varepsilon_{rr}^2 + \varepsilon_{zz}^2) \quad (S10)$$

Eq. S10 can then be simplified by factoring to arrive at Eq. 5:

$$u = \frac{E}{2(1+\nu)}\left[\frac{\nu(\varepsilon_{zz} + 2\varepsilon_{rr})^2}{(1-2\nu)} + \varepsilon_{zz}^2 + 2\varepsilon_{rr}^2\right] \quad (5)$$

Strain values in Eq. 5 are calculated from the measured quantities $r_{avg}$, $z_{avg}$, Δr, and Δz using our minimal cylinder analysis (Methods 2.1.3).

We next focus on the acoustic cylindrical mode of motion that describes compression in the z-axis along with radial extension (and *vice versa*). The average potential energy of this mode can be estimated from the equipartition theorem, $U = \frac{1}{2}k_b T$, where $k_b$ is the Boltzmann constant and T is the simulation temperature, 300 K. We then integrate Eq. 5 over the body volume, Eq. 1, and with the above-mentioned energy from equipartition theorem we solve for Young's modulus:

$$E = \frac{k_b T(1-\nu-2\nu^2)}{V(\varepsilon_{zz}^2 - \nu\varepsilon_{zz}^2 + 2\varepsilon_{rr}^2 + 4\nu\varepsilon_{zz}\varepsilon_{rr})} \cdot (6)$$

**S5.** *The heterotypic nucleosome has an exposed CENP-A RG loop*

The symmetry breaking in the heterotypic nucleosome across the pseudo-dyad, also propagates away from the CENP-A:H3.3 interface to the RG loop (R80, G81) of CENP-A, L1 (Fig. S1C). The CENP-A histone displays increased exposure of L1 to solvent in the case of the heterotypic nucleosome in comparison to the CENP-A nucleosome (Movie S1). For viable kinetochore formation at the heterotypic nucleosome, other proteins such as CENP-N must be recruited (7, 8). The high degree of solvent exposure in the heterotypic nucleosome may indicate that CENP-N is able to bind to its established binding site, CENP-A R80 and G81 (8). It is not yet understood if a single copy of CENP-C and CENP-N are sufficient for kinetochore formation, though minimally, our work on the rigidification of CENP-A when bound to its partners (Melters, Pitman, Rakshit *et al. in press 2019*), indicates that these factors would further stabilize CENP-A of heterotypic nucleosomes at ectopic sights.

**Supporting References**

1. Winogradoff, D., H. Zhao, Y. Dalal, and G.A. Papoian. 2015. Shearing of the CENP-A